\documentclass{article}
\usepackage{spconf,amsmath,graphicx,hyperref}
\usepackage{framed,multirow}
\usepackage{booktabs}

\title{HEAR: Hierarchically Enhanced Aesthetic Representations for Multidimensional Music Evaluation}

\name{
Shuyang Liu$^{12}$$^{\ast}$ \qquad Yuan Jin$^{12}$$^{\ast}$ \qquad Rui Lin$^{12}$ \qquad Shizhe Chen$^3$ \qquad Junyu Dai$^{12}$$^{\dagger}$ \qquad Tao Jiang$^{12}$
\thanks{
        $^{\ast}$ Equal contribution. 
        $^{\dagger}$ Project Lead. 
    }}

\address{
{$^1$ $\epsilon$ar-LAB} \\
{$^2$ ZiYouLiangJi (Shanghai) Information Technology Co., Ltd}\\
{$^3$ Shanghai Conservatory of Music}}

\begin{document}

\maketitle

\begin{abstract}
Evaluating song aesthetics is challenging due to the multidimensional nature of musical perception and the scarcity of labeled data. We propose HEAR, a robust music aesthetic evaluation framework that combines: (1) a multi-source multi-scale representations module to obtain complementary segment- and track-level features, (2) a hierarchical augmentation strategy to mitigate overfitting, and (3) a hybrid training objective that integrates regression and ranking losses for accurate scoring and reliable top-tier song identification. Experiments demonstrate that HEAR consistently outperforms the baseline across all metrics on both tracks of the ICASSP~2026 SongEval benchmark. The code and trained model weights are available at \url{https://github.com/Eps-Acoustic-Revolution-Lab/EAR_HEAR}.

\end{abstract}

\begin{keywords}
Music Aesthetics Evaluation, Audio Representation Learning, Data Augmentation, SongEval
\end{keywords}


\section{Introduction}
\label{sec:intro}
With the rapid development of generative music models, automated music aesthetic evaluation has become increasingly important, yet remains challenging. Existing approaches, such as Audiobox-Aesthetics~\cite{tjandra2025meta}, employ simple Transformer-based architectures to predict multidimensional aesthetic scores but struggle to capture rich musical characteristics. While SongEval~\cite{yao2025songeval} establishes a high-quality benchmark, its limited data scale challenges the training of robust aesthetic evaluators. 
To this end, we propose a robust framework with the following main contributions:
\begin{itemize}
\setlength{\itemsep}{2pt}
\setlength{\parskip}{0pt}
\setlength{\parsep}{0pt}
\item We propose \textbf{HEAR}, which synergizes a multi-source multi-scale representations module and a hierarchical augmentation strategy to capture robust musical features under limited labeled data.
\item We introduce a hybrid training objective to enable accurate aesthetic scoring and top-tier song identification, achieving significant improvements over baselines on the ICASSP 2026 SongEval benchmark.
\end{itemize}

\begin{figure}[t] 
    \centering
    \centerline{\includegraphics[width=0.5\textwidth]{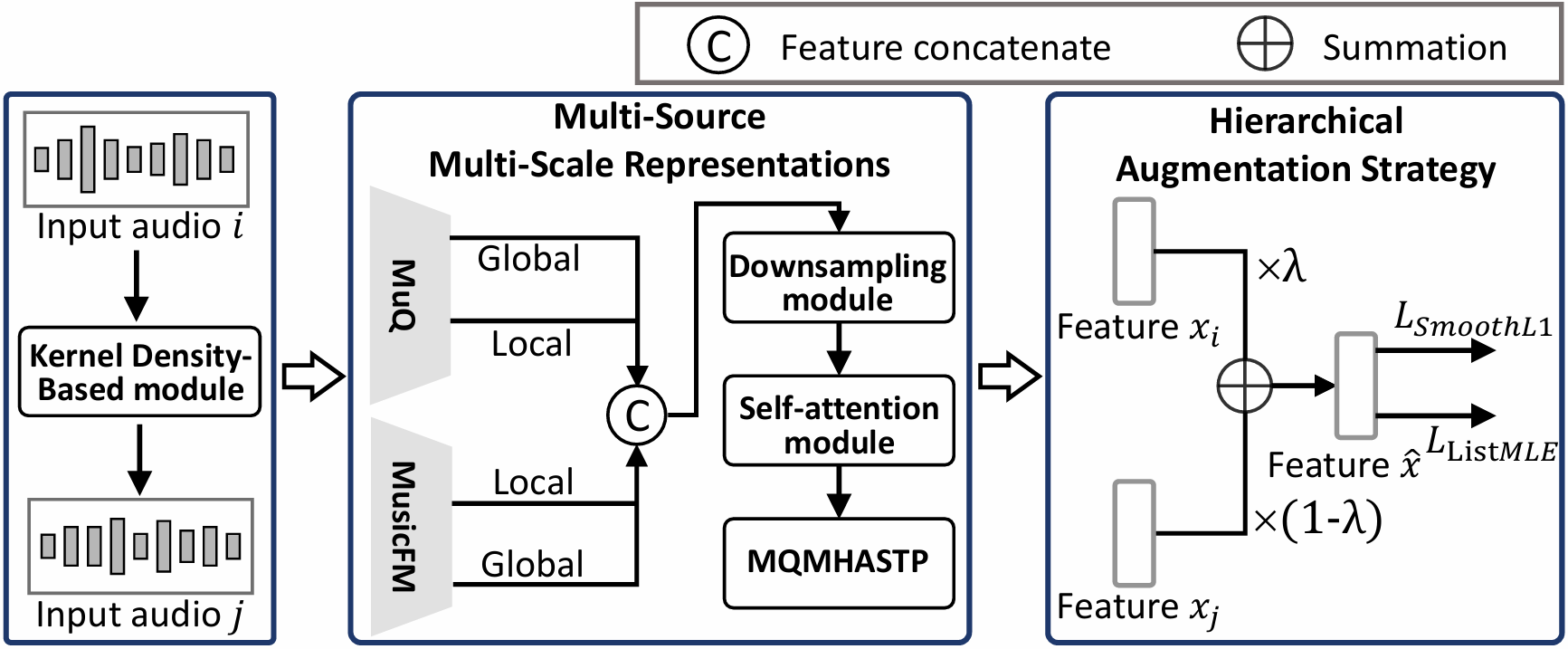}}
    \vspace{-5pt}
    \caption{Overview of our proposed HEAR.} 
    \label{fig:main_framework_figure}
  \vspace{-7pt} 
\end{figure}

\section{Method}
\subsection{Overview}
Figure~1 illustrates the overall architecture of the proposed HEAR.
It consists of three key components: (1) multi-source multi-scale representations module for comprehensive audio
features capturing, (2) hierarchical augmentation strategy at both
data and feature levels, and (3) a hybrid training objective
for multidimensional aesthetic prediction. These components constitute a robust and effective framework for music aesthetic evaluation.

\subsection{Multi-Source Multi-Scale Representations}
Inspired by Songformer \cite{hao2025songformer}, we employ both MuQ~\cite{zhu2025muq} and MusicFM~\cite{won2024foundation} to extract complementary local segment-level and global
track-level multi-scale music representations, followed by downsampling, self-attention, and a 
Multi-Query Multi-Head Attention Statistical Pooling (MQMHASTP) module 
\cite{zhao2022multi}, which enables the model to capture how temporal, spectral, harmonic, and content cues contribute to different aesthetic dimensions while converting variable-length features into fixed-length representations.

\subsection{Hierarchical Augmentation Strategy}
We introduce a Hierarchical Augmentation strategy that operates at both data and feature levels. 
At the data level, we apply a conservative audio augmentation pipeline to expand the training set, 
with details summarized in Section~\ref{sec:data_augmentation}.
At the feature level, we employ C-Mixup \cite{yao2022c}, which performs
conditional mixing by sampling neighboring examples with higher
probability in the label space via Kernel Density Estimation
(KDE):
\begin{equation}
P\big((x_j, y_j) \mid (x_i, y_i)\big)
\propto \exp\Big(-\frac{d(i,j)^2}{2\sigma^2}\Big),
\end{equation}
where $d(i,j)$ denotes the Euclidean distance between
the labels of two examples. A mixed
feature–label pair is then obtained by convex combination:
\begin{equation}
\hat{x} = \lambda x_i + (1-\lambda)x_j, \quad
\hat{y} = \lambda y_i + (1-\lambda)y_j,
\end{equation}
with $\lambda$ sampled from a Beta distribution, i.e., $\lambda \sim \mathrm{Beta}(\alpha, \alpha)$.

\subsection{Hybrid Training Objective}
To jointly support aesthetic score prediction and top-tier song identification, we adopt a hybrid objective $L_{total}$ combining SmoothL1 loss \cite{girshick2015fast} for regression and the listwise ranking loss
ListMLE \cite{xia2008listwise} for modeling relative ordering among
samples:
\begin{equation}
L_{total} = L_{SmoothL1} + \beta L_{ListMLE},
\end{equation}
where $\beta$ weights the ranking term to mitigate sensitivity to unreliable orderings from subjective scores, especially among similar samples.


\begin{table}[!b]
\vspace{-20pt}
\centering
\caption{Data-level augmentation settings used for training.}
\label{tab:augmentation}
\small
\setlength{\tabcolsep}{4pt}
\renewcommand{\arraystretch}{1.1}
\begin{tabular}{p{3.2cm} p{4.8cm}}
\hline
\textbf{Augmentation} & \textbf{Parameters (Probability)} \\
\hline
Time-stretch & 0.99--1.01 ($p=0.4$) \\
Time shift & $\pm$0.5 s ($p=0.4$) \\
Pitch shift & $\pm$10 cents ($p=0.5$) \\
Gain perturbation & $-2$ to $+2$ dB ($p=0.9$) \\
High-pass filter & 80--120 Hz ($p=0.3$) \\
Low-pass filter & 15--18 kHz ($p=0.3$) \\
Parametric EQ & 7 bands, $-3$ to $+3$ dB ($p=0.3$) \\
Gaussian noise & 30--50 dB SNR ($p=0.25$) \\
\hline
\end{tabular}
\end{table}

\section{Experiments and Analysis}
\subsection{Experiment Setup}
\subsubsection{Implementation Details}
\label{sec:data_augmentation}

All experiments are conducted on the SongEval dataset
\cite{yao2025songeval}, which contains 2,399 songs annotated across five
aesthetic dimensions. Following the official protocol, 200 samples are
used for validation, while the remaining data are augmented using eight data-level strategies summarized in Table~\ref{tab:augmentation} for training. 
The model is trained using the Adam optimizer with a learning rate of $1\times10^{-5}$ and a weight decay of $1\times10^{-3}$ with a batch size of 8. 
The hyperparameter $\beta$, weighting the ranking objective, is set to $0.15$ for Track~1 and $0.05$ for Track~2. For C-Mixup, the bandwidth parameter $\sigma$ of the Gaussian kernel is
set to 1, and $\alpha$ in Beta distribution is set to 2.

\subsubsection{Evaluation Metrics}
We evaluate the model using four official metrics: the Linear Correlation Coefficient (LCC) ~\cite{sedgwick2012pearson} measures the linear alignment between predicted and ground-truth scores, while Spearman’s Rank Correlation Coefficient (SRCC)~\cite{sedgwick2014spearman} and Kendall’s Tau Rank Correlation (KTAU)~\cite{mcleod2005kendall} assess ranking consistency. Top-Tier Accuracy (TTA) measures top-tier song identification via an F1 score
with official thresholding.


\begin{table}[!t]
\centering
\vspace{-15pt}
\caption{Effects of different modules in our method. Baseline denotes the officially provided model trained on SongEval.}
\vspace{0.5em}
\label{tab:ablation_modules}
\renewcommand{\arraystretch}{0.8}
\setlength{\tabcolsep}{3pt}
\scalebox{0.8}{
\begin{tabular}{c c c c c c c c c}
\toprule[2pt]
Track & Baseline & MMR & HAS & $L_{total}$ 
& LCC $\uparrow$ & SRCC $\uparrow$ & KTAU $\uparrow$ & TTA $\uparrow$ \\ 
\midrule[1pt]
\multirow{8}{*}{1}  
& $\surd$ &  &  &  &0.913  &0.900  &0.745  &0.818  \\
& $\surd$ & $\surd$ &  &  & 0.914 & 0.907 & 0.746 & 0.845 \\
& $\surd$ &  &$\surd$  &  &0.914  & 0.900 &0.743  &0.841  \\
& $\surd$ &  &  &$\surd$  & 0.911 &0.904  & 0.748 & 0.841 \\
& $\surd$ & $\surd$ & $\surd$ & & 0.914 & 0.902 &0.742  & 0.853  \\
& $\surd$ & $\surd$ &  &$\surd$  &0.917  &0.907  &0.741  &0.865  \\
& $\surd$ &  & $\surd$ &$\surd$  & 0.916 & 0.904 & 0.745 & 0.859 \\
& $\surd$ & $\surd$ & $\surd$ & $\surd$ 
& \textbf{0.920} & \textbf{0.910} & \textbf{0.752} & \textbf{0.871} \\
\midrule[1pt]
\multirow{8}{*}{2}
& $\surd$ &  &  &  & 0.908 & 0.897 & 0.736 & 0.826 \\
& $\surd$ & $\surd$ &  &  & 0.909 & 0.899 & 0.733 & 0.840 \\
& $\surd$ &  &$\surd$  &  & 0.906  & 0.896 &0.737  & 0.832 \\
& $\surd$ &  &  &$\surd$  &0.902  & 0.892 & 0.728 & 0.838 \\
& $\surd$ & $\surd$ & $\surd$ &  & 0.909 & 0.897 & 0.733 & 0.843 \\
& $\surd$ & $\surd$ &  & $\surd$ & 0.910 & 0.899 & 0.737 & 0.846 \\
& $\surd$ &  & $\surd$ & $\surd$ &  0.902 & 0.896 & \textbf{0.739} & 0.840 \\
& $\surd$ & $\surd$ & $\surd$ & $\surd$ 
& \textbf{0.912} & \textbf{0.900} & 0.737 & \textbf{0.853} \\
\bottomrule[2pt]
\end{tabular}
}
\vspace{-10pt}
\end{table}

\subsection{Overall Performance}

As shown in Table \ref{tab:ablation_modules}, all experiments follow the official settings\footnote{\url{https://aslp-lab.github.io/Automatic-Song-Aesthetics-Evaluation-Challenge/}}, with Track~1 and Track~2 corresponding to single- and multidimensional aesthetic prediction, respectively. Incrementally adding MMR (Multi-Source Multi-Scale Representations), HAS (Hierarchical Augmentation Strategy), and the hybrid training loss $L_{total}$ to the baseline yields consistent performance gains. The gains on both tracks validate the effectiveness of HEAR for music aesthetic evaluation and top-tier song identification, further evidenced by the official rankings of 2nd/19 on Track~1 and 5th/17 on Track~2 in the Automatic Song Aesthetics Evaluation Challenge.

\section{Conclusion}

We presented a robust framework HEAR for multidimensional music aesthetic evaluation, which effectively addresses the challenges of limited labeled data and complex musical perception. Our method synergistically combines multi-source multi-scale representations, a hierarchical augmentation strategy, and a hybrid training objective. Experiments on the ICASSP 2026 SongEval benchmark confirm that our approach consistently surpasses baseline methods, achieving superior performance in both aesthetic scoring and top-tier song identification. 

\newpage
\bibliographystyle{IEEEbib}
\bibliography{strings,refs}

\end{document}